\documentclass[12pt]{article}

\catcode`@=12
\begin{document}
\thispagestyle{empty}
\begin{flushright} 
UCRHEP-T329\\ 
January 2002\
\end{flushright}
\vspace{0.5in}
\begin{center}
{\LARGE	\bf Pattern of the Approximate Mass\\Degeneracy of Majorana 
Neutrinos\\}
\vspace{1.5in}
{\bf Ernest Ma\\}
\vspace{0.2in}
{\sl Physics Department, University of California, Riverside, 
California 92521\\}
\vspace{1.5in}
\end{center}
\begin{abstract}\
In view of the recently reported evidence for a nonzero Majorana mass of the 
electron neutrino, together with the established phenomena of atmospheric and 
solar neutrino oscillations, the case of three nearly mass-degenerate 
Majorana neutrinos is now a distinct possibility.  I show in this 
paper how a natural pattern of symmetry breaking in the recently proposed 
$A_4$ model of Majorana neutrino masses can accommodate the data on neutrino 
oscillations, resulting in the predictions $\sin^2 2 \theta_{atm} = 1$ and 
$\sin^2 2 \theta_{sol} = 2/3$.
\end{abstract}

\newpage
\baselineskip 24pt

In the past several years, there has been mounting evidence for neutrino 
oscillations \cite{atm,solar,lsnd}.  Since they require only neutrino mass 
differences, the possibility of nearly degenerate neutrino masses is often 
considered \cite{amr}.  Recently, the first positive evidence for 
neutrinoless double beta decay has been reported \cite{klapdor} which 
may be interpreted as an effective nonzero Majorana mass for the electron 
neutrino. Combined with the atmospheric and solar neutrino data, there is 
now a plausible argument for three nearly mass-degenerate Majorana neutrinos 
\cite{klsa}.  However, the charged-lepton 
masses are certainly not degenerate, so whatever symmetry is used to maintain 
the neutrino mass degeneracy must be broken.  To implement this idea in a 
renormalizable field theory, the discrete symmetry $A_4$ was proposed 
recently \cite{mara} where its spontaneous breaking results in charged-lepton 
masses and its explicit soft breaking results in neutrino mass differences.

The proposed $A_4$ model is based on a simple model of neutrino masses 
\cite{ma01}, where a leptonic Higgs doublet $\eta = (\eta^+,\eta^0)$ and 
three right-handed neutral singlet fermions $N_{iR}$ are added to the 
minimal standard model of particle interactions.  These new particles 
may all be at or below the TeV energy scale, so that the seesaw mechanism 
\cite{seesaw} may be tested experimentally at future accelerators.  With 
the report of a possible discrepancy in the experimental measurement 
\cite{g-2} of the muon anomalous magnetic moment with the theoretical 
prediction, this leptonic Higgs model was used \cite{mr} to constrain 
the masses and couplings of these new particles.  However, a sign error has 
been discovered \cite{sign} in the theoretical calculation, hence the 
experimental discrepancy is now only $1.6\sigma$, which is not much of a 
constraint on this model.

In this paper the explicit soft breaking of the $A_4$ symmetry is shown to 
allow for a natural solution with the predictions $\sin^2 2\theta_{atm} = 1$ 
and $\sin^2 2\theta_{sol} = 2/3$ which agree well with present data and 
may be tested more precisely in future neutrino-oscillation experiments. 
At the same time, the new particles $N_{iR}$ as well as an assortment of Higgs 
bosons with specific properties \cite{mara} are predicted to be accessible 
at future high-energy accelerators and their decays into leptons will map 
out the neutrino mass matrix.

Under $A_4$ and $L$ (lepton number), the color-singlet fermions and scalars 
of this model transform as follows.
\begin{eqnarray}
(\nu_i, l_i)_L &\sim& (\underline {3}, 1), \\ 
l_{1R} &\sim& (\underline {1}, 1), \\ 
l_{2R} &\sim& (\underline {1}', 1), \\ 
l_{3R} &\sim& (\underline {1}'', 1), \\
N_{iR} &\sim& (\underline {3}, 0), \\ 
\Phi_i = (\phi_i^+, \phi_i^0) &\sim& (\underline {3}, 0), \\ 
\eta = (\eta^+, \eta^0) &\sim& (\underline {1}, -1).
\end{eqnarray}
Hence its Lagrangian has the invariant terms
\begin{equation}
{1 \over 2} M N_{iR}^2 + f \bar N_{iR} (\nu_{iL} \eta^0 - l_{iL} \eta^+) + 
h_{ijk} \overline {(\nu_i,l_i)}_L l_{jR} \Phi_k + h.c.,
\end{equation}
where
\begin{equation}
h_{i1k} = h_1 \left[ \begin{array} {c@{\quad}c@{\quad}c} 1 & 0 & 0 \\ 
0 & 1 & 0 \\ 0 & 0 & 1 \end{array} \right], ~~ h_{i2k} = h_2 \left[ 
\begin{array} {c@{\quad}c@{\quad}c} 1 & 0 & 0 \\ 0 & \omega & 0 \\ 
0 & 0 & \omega^2 \end{array} \right], ~~ h_{i3k} = h_3 \left[ \begin{array} 
{c@{\quad}c@{\quad}c} 1 & 0 & 0 \\ 0 & \omega^2 & 0 \\ 0 & 0 & \omega 
\end{array} \right],
\end{equation}
with $\omega^3 = 1$. Thus the neutrino mass matrix (in this basis) is 
proportional to the unit matrix with magnitude $f^2 u^2/M$, where $u = 
\langle \eta^0 \rangle$, whereas the charged-lepton mass matrix is given by
\begin{equation}
{\cal M}_l = \left[ \begin{array} {c@{\quad}c@{\quad}c} h_1 v_1 & h_2 v_1 & 
h_3 v_1 \\ h_1 v_2 & h_2 \omega v_2 & h_3 \omega^2 v_2 \\ h_1 v_3 & h_2 
\omega^2 v_3 & h_3 \omega v_3 \end{array} \right],
\end{equation}
where $v_i = \langle \phi_i^0 \rangle$.  Now rotate ${\cal M}_l$ on the left by
\begin{equation}
U_L^\dagger = {1 \over \sqrt 3} \left[ \begin{array} {c@{\quad}c@{\quad}c} 1 
& 1 & 1 \\ 1 & \omega^2 & \omega \\ 1 & \omega & \omega^2 \end{array} \right],
\end{equation}
then
\begin{equation}
U_L^\dagger {\cal M}_l = \left[ \begin{array}{c@{\quad}c@{\quad}c} v & v' & 
v'' \\ v'' & v & v' \\ v' & v'' & v \end{array} \right] \left[ \begin{array} 
{c@{\quad}c@{\quad}c} h_1 & 0 & 0 \\ 0 & h_2 & 0 \\ 0 & 0 & h_3 \end{array} 
\right],
\end{equation}
where
\begin{eqnarray}
v &=& {1 \over \sqrt 3} (v_1 + v_2 + v_3), \\ 
v' &=& {1 \over \sqrt 3} (v_1 + \omega v_2 + \omega^2 v_3), \\ 
v'' &=& {1 \over \sqrt 3} (v_1 + \omega^2 v_2 + \omega v_3).
\end{eqnarray}

As shown in Ref.~[7], $v_1=v_2=v_3$ is a natural solution of the 
$A_4$-symmetric Higgs potential, in which case $v'=v''=0$ and $U_L^\dagger 
{\cal M}_l$ is diagonal.  Hence
\begin{equation}
{\cal M}_\nu = {f^2 u^2 \over M} U_L^T U_L = {f^2 u^2 \over M} \left[ 
\begin{array} {c@{\quad}c@{\quad}c} 1 & 0 & 0 \\ 0 & 0 & 1 \\ 
0 & 1 & 0 \end{array} \right]
\end{equation}
in the $(\nu_e,\nu_\mu,\nu_\tau)$ basis. This shows that $\nu_\mu$ mixes 
maximally with $\nu_\tau$, but since all physical neutrino masses are 
degenerate, there are no neutrino oscillations.  To break the degeneracy, 
let ${\cal M}_{ij} = M \delta _{ij} + m_{ij}$, then
\begin{equation}
({\cal M}^{-1})_{ij} \simeq M^{-1} \delta_{ij} - M^{-2} m_{ij}.
\end{equation}
Whereas $m_{ij}$ is assumed arbitrary in Ref.~[7], it is required here to 
be invariant under $U_L$, i.e.
\begin{equation}
U_L^T m_{ij} U_L = m_{ij}.
\end{equation}
It is then a simple exercise to show that the most general solution is of 
the form
\begin{equation}
m_{ij} = \left[ \begin{array} {c@{\quad}c@{\quad}c} 2\delta + 2\delta' & 
\delta' & \delta' \\ \delta' & \delta & \delta \\ \delta' & \delta & \delta 
\end{array} \right].
\end{equation}

Consider first the case $\delta'=0$, then
\begin{equation}
{\cal M}_\nu \simeq {f^2 u^2 \over M} \left[ \begin{array} 
{c@{\quad}c@{\quad}c} 1-2\delta/M & 0 & 0 \\ 0 & -\delta/M & 1-\delta/M \\ 
0 & 1-\delta/M & -\delta/M \end{array} \right],
\end{equation}
which has eigenvalues proportional to $1-2\delta/M$, $1-2\delta/M$, and $-1$, 
corresponding to the eigenstates $\nu_e$, $(\nu_\mu+\nu_\tau)/\sqrt 2$, and 
$(\nu_\mu-\nu_\tau)/\sqrt 2$ respectively.  This shows that the threefold 
degeneracy of ${\cal M}_\nu$ is broken by $\delta$ to a twofold degeneracy 
with $\nu_\mu-\nu_\tau$ maximal mixing and $\Delta m^2 \simeq 4 \delta f^4 
u^4/M^3$, which is desirable for explaining atmospheric neutrino oscillations 
\cite{atm}.  It also provides a natural reason for having $\delta' << \delta$ 
because $\delta'$ breaks even the twofold degeneracy, as discussed below.

To see how $\delta' \neq 0$ affects $m_{ij}$ of Eq.~(19), rotate 
${\cal M}_\nu$ of Eq.~(20) to the basis spanned by $\nu_e$, $(\nu_\mu + 
\nu_\tau)/\sqrt 2$, and $(\nu_\mu - \nu_\tau)/\sqrt 2$.  Then
\begin{equation}
{\cal M}_\nu \simeq {f^2 u^2 \over M} \left[ \begin{array} 
{c@{\quad}c@{\quad}c} 1-2\delta/M-2\delta'/M & -\sqrt 2 \delta'/M & 0 \\ 
-\sqrt 2 \delta'/M & 1-2\delta/M & 0 \\ 0 & 0 & -1 \end{array} \right],
\end{equation}
which has the solution
\begin{equation}
\left[ \begin{array} {c} \nu_1 \\ \nu_2 \\ \nu_3 \end{array} \right] = \left[ 
\begin{array} {c@{\quad}c@{\quad}c} \cos \theta & \sin \theta/\sqrt 2 & 
\sin \theta/\sqrt 2 \\ -\sin \theta & \cos \theta/\sqrt 2 & \cos \theta 
/\sqrt 2 \\ 0 & 1/\sqrt 2 & -1/\sqrt 2 \end{array} \right] \left[ 
\begin{array} {c} \nu_e \\ \nu_\mu \\ \nu_\tau \end{array} \right],
\end{equation}
where $\tan \theta = (\sqrt 3 - 1)/\sqrt 2$, and
\begin{eqnarray}
m_1 &\simeq& {f^2 u^2 \over M} \left[ 1 - {2\delta \over M} - {(\sqrt 3 + 1) 
\delta' \over M} \right], \\ 
m_2 &\simeq& {f^2 u^2 \over M} \left[ 1 - {2\delta \over M} + {(\sqrt 3 - 1) 
\delta' \over M} \right],
\end{eqnarray}
and $m_3 \simeq -f^2 u^2/M$.  Hence
\begin{equation}
\Delta m_{12}^2 \simeq {4 \sqrt 3 \delta' f^4 u^4 \over M^3}, ~~~ 
\sin^2 2 \theta_{12} = {2 \over 3}.
\end{equation}
This is then a satisfactory explanation of the solar neutrino data 
\cite{solar} with a large mixing angle.  Numerically, let the common mass 
of all three neutrinos be $f^2 u^2/M = 0.4$ eV \cite{klapdor}, and 
$\delta/M = 3.9 \times 10^{-3}$, $\delta'/M = 3.6 \times 10^{-5}$; then 
$(\Delta m^2)_{atm} = 2.5 \times 10^{-3}$ eV$^2$ and 
$(\Delta m^2)_{sol} = 4.0 \times 10^{-5}$ eV$^2$, in good agreement with 
present data.  Note also that these numbers are not spoiled by radiative 
corrections \cite{klsa}.

As Eq.~(22) shows, under the assumption of $v_1=v_2=v_3$ and that of Eq.~(18), 
the electron neutrino $\nu_e$ has only $\nu_1$ and $\nu_2$ components, i.e. 
$U_{e3} = 0$.  This is perfectly consistent with present data.  However, if 
$U_{e3}$ is indeed zero, then there can be no $CP$-violating effects in 
neutrino oscillations.  In the context of the present model, if the Higgs 
potential has soft terms which break the $A_4$ symmetry, then $v'$ and $v''$ 
of Eqs.~(14) and (15) will not be zero, but may be assumed to be small. 
In that case, $(h_1,h_2,h_3)$ of Eq.~(12) are still approximately proportional 
to $(m_e,m_\mu,m_\tau)$, and it is easy to show that the rotation due to 
$v' \neq 0$ and $v'' \neq 0$ results in $|U_{e3}| \simeq |v'/v\sqrt 2|$, 
which is bounded by reactor experiments \cite{react} to be less than about 
0.16.

The heavy right-handed singlet fermions $N_{iR}$ are of course also nearly 
degenerate in mass.  As discussed in Ref.~[8], they will decay into charged 
leptons plus either a $W^\pm$ boson or a charged Higgs boson.  The mass 
eigenstates of $N_{iR}$ are given by $N_1 \cos \theta + (N_2 + N_3) 
\sin \theta/\sqrt 2$, $-N_1 \sin \theta + (N_2 + N_3) \cos \theta/\sqrt 2$, 
and $(N_2 - N_3)/\sqrt 2$, with masses $M + 2 \delta + (\sqrt 3 + 1) \delta'$, 
$M + 2 \delta - (\sqrt 3 - 1) \delta'$, and $M$ respectively.  Their couplings 
are given by Eqs.~(8) and (11), i.e. $N_1 \to (e + \mu + \tau)/\sqrt 3$, 
$N_2 \to (e + \omega \mu + \omega^2 \tau)/\sqrt 3$, and $N_3 \to (e + 
\omega^2 \mu + \omega \tau)/\sqrt 3$.  An equal and incoherent mixture of 
all three $N$'s will of course decay equally into $e$, $\mu$, and $\tau$.

In conclusion, the case of three nearly mass-degenerate Majorana neutrinos 
in a renormalizable field theory based on the discrete symmetry $A_4$ is 
studied and found to accommodate a natural solution with two mass splittings, 
one larger than the other because it breaks the threefold degeneracy only 
down to a twofold degeneracy.  This pattern is ideal for understanding the 
recently reported evidence for a nonzero effective Majorana mass of the 
electron neutrino, and the established phenomena of atmospheric and solar 
neutrino oscillations.  It predicts $\sin^2 2 \theta_{atm} = 1$ and 
$\sin^2 2 \theta_{sol} = 2/3$ with a zero or small $U_{e3}$.  It also 
predicts particles at the TeV energy scale which are responsible for 
the proposed pattern.  As such, future neutrino-oscillation experiments are 
complementary to future high-energy accelerator experiments in the 
unambiguous test of this model.

This work was supported in part by the U.~S.~Department of Energy
under Grant No.~DE-FG03-94ER40837.

\bibliographystyle{unsrt}

\end{document}